# A Method for Counting, Tracking and Monitoring of Visitors with RFID sensors

## Model of Study: M. Hatzidakis residence

A.Gazis, K.Stamatis, E.Katsiri

Department of Electrical and Computer Engineering
Polytechnic School, Democritus University of Thrace
Xanthi 67100, Greece
agazis@ee.duth.gr, kstamati@ee.duth.gr, ekatsiri@ee.duth.gr

*Abstract* — **This publication presents a method responsible for counting tracking and monitoring visitors inside a building. The site examined is Manos Hatzidakis' House, situated in Xanthi. Specifically, we have conducted a study, which provides recommendations, regarding the installation of sensors in the building. We also present the communication protocols of the computer network used in order to ensure the efficient communication between the space examined and the sensor network. Finally, we describe the process of creating a website, which is designed to store and view the data.**

*Keywords* — *counting, monitoring, tracking, RFID, Raspberry Pi, wireless sensor network, django python framework*

## I. INTRODUCTION-PRESENTATION OF THE PRESERVED AREA

The main goal of this paper is to propose a cost effective and reliable method in order to accurately count, track and monitor visitors. The model of our study was based in closed area structures using RFID and Wireless Sensor Networks. More specifically, the building we chose as our model of study to be monitored is Hatzidaki's house. The city of Xanthi is divided in two sections: the urban sector and the Old Town. The Old Town is an area of 380,000 m², in the northern part of Xanthi and its construction dates back to 1829. One of the most important monuments of this area is the historical building "Xanthi's Guard", which was built in 1895, by Isaac Daniès. It is believed that the building was constructed by an Austrian architect, of unknown identity. On April 17, 2016, it was restored and renovated, under the supervision of the Region of Eastern Macedonia and Thrace. The site is a monument with neo-classical elements and Baroque style, which consists of three floors and covers a plot of $1.317 \text{ m}^2$ [1] as presented in Figure [1].

In this publication, the ground floor as well as the first floor will be studied, so as to build a wireless sensor network that will track and monitor visitors. These metrics will be sent and stored on a site, which processes and displays the clues, as real-time data, which are analyzed and not provided in a raw format. The type of sensors, as well as the way of their communication will be extensively analyzed in the following section of this study. The model is consisted of four rooms, three of which are available to the public. Each room has an entrance, where an electronic count meter is installed, which is able to record data, which are related to the entry of each visitor. In the fourth room, which is not available to the public, there is a computer, which performs the process of controlling and coordinating the others, through their local interconnection.

## II. AIMS AND OBJECTIVES

In the following section, we present the basic structure of the concepts of work selected. These include our initial approach in order to study and select the right technology for the sensors and other necessary electronic devices. Moreover, we present the format data used to analyze the information of the sensors indications and the application proposed in order to visualize them. We will present our initial approach to solving the problem of counting tracking and monitoring of visitors and our proposed solution.

This paper aims to address the evaluation of a low-cost technique, examining whether it can provide reliable monitoring, tracking and counting of a prefixed point of entrance (one for each floor). In order to achieve this, a pilot wireless sensor network (WSN) was developed with the following aim:

1. Count and visualize all available people in an area

2. Track the visitors of the building via the data produced by RFID tags

3. Monitor the changes to the overall system and number of people in a building.

This work has as an objective:

1. Design a WSN responsible for detecting and automatically counting the number of people.

2. Present the custom WSN architecture used to visualize and store the data.

3. Suggest the usage of the available data as an enhancement of the current evacuation techniques.

## A. Initial Approach

As previously mentioned we have chosen as our model of study M. Hatzidakis' House. Firstly, we studied the plans regarding each floor in order to understand whether the idea proposed could be implemented in this model of study. The problem concerned a large multistorey building and we concentrated on the communication of the sensors in different floors and the points where we would place our sensors. We reached to the conclusion that a possible solution could be provided, since the building did not have multiple entrances and exits. Thus, a number of ways could be developed, in order to monitor these points. Firstly, we had to count and track the number of visitors in each room and throughout the building, at regular intervals. One of the main problems was the location of the sensors, as well as the technology, which was used. The model of study dictated that we had to develop a low-cost system, which would be efficient and characterized by high response rates both for the data transfer as well as the page refresh rate. As far as the installation is concerned, we extensively studied the electrical designs of each floor plan, so as to ensure that no electromagnetic interruptions or any kind of noise close to the input/output sensors' signals would be produced. We also aimed for the continuous and adequate supply of power to the sensors, as well as the reception of their indications.

Specifically, we concluded that every visitor will be given a separate RFID tag on the ground floor, since this is the only entrance to the building. The RFID tag is the visitor's identity for our system. This tag will track the person, while moving in the rooms of the house. In order to read the tags of each visitor, appropriate RFID readers and the necessary RFID antennas will be used in each room under surveillance. In particular, each room will be equipped with an RFID reader to read the tags of the visitors. The RFID reader will be connected to a Raspberry Pi motherboard. The Raspberry Pi motherboard will receive the data from the reader and will forward it to our system, via the internet, while pre-processing the data.

The data of the RFID reader will be of the following format:

<TagId | Name>=<01008C7200 | Visitor Name1, 01008C7201 | Visitor Name2, 01008C7202 | Visitor Name3, … >

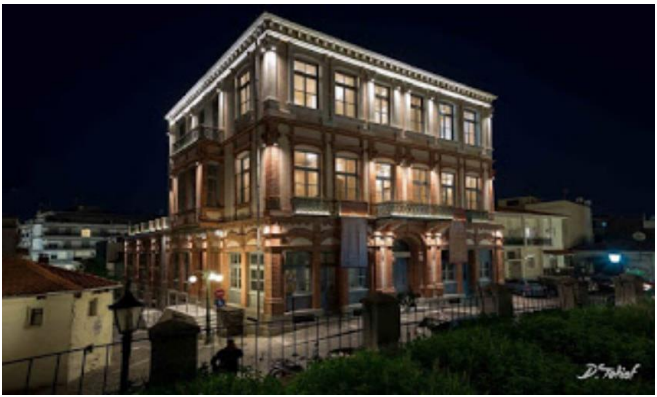

Fig. 1. M.Hatzidakis residence [2] .

## B. Proposed Solution

We recommend the use of Passive High Frequency RFID Tags, with an operating frequency of 13.56 MHz [3]. With this type of sensor, a maximum reading distance of 2 meters is possible, and a reader can be used in conjunction with a customized antenna to extend the reading range for longer distances [4].

We also recommend the Raspberry Pi 3 Model B [5] platform, due to its ease of use, the plethora of tools available, its low cost and its ease of interconnection with other devices of our system, as well as the internet.

## C. Floor Plans of the Building Examined

This section provides the floor plans of M. Hatzidakis' house, which was the basic model for the development of the application. The following plans were extensively studied, so as to extract the application, select the sensors' positions and determine the most efficient system. These illustrations picture the electrical designs of the ground, as well as the first floor of the establishment studied as presented in Figure [2] and [3] respectively.

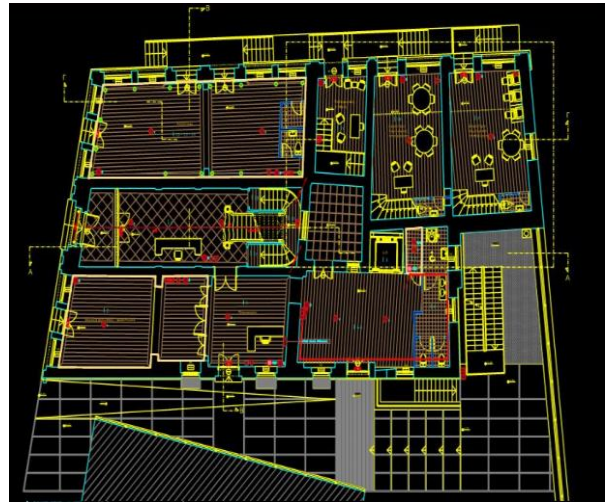

Fig. 2. Floor plan of the ground floor.

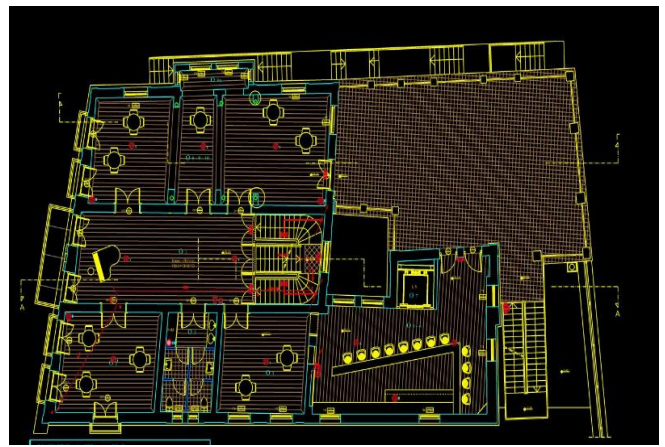

Fig. 3. Floor plan of the first floor.

### D. Communication Protocols and Visitor Contribution

The proposed solution consists of two equally important parts. In particular, a system must be created that will consist of a basic and an online web application, which will enable the user to log all the control points and the labels. Specifically, the back-end application is responsible for reading the labels and sending the data to the online web application, which is linked to a database. Firstly, the user-manager must have access to the online application, so as to enter the checkpoints and labels. When registering checkpoints, the user provides the IP address or MAC address of each Raspberry Pi, as well as its physical location (e.g. room_1). Additionally, the user enters the tag and the information of the visitor. Furthermore, the online application provides visitor visibility and it can be accessed by the manager, in order to check the last known position of the visitors of M. Hatzidakis House, as well as the last time stamp reading [6].

In this project, Raspberry Pi will run in headless mode, which means that there is no user interface. This minicomputer will gather the HF RFID tag values and assist the Background Data Collection Application. The WSN consists of the RFID readers [7] which are connected to a Raspberry Pi which serves as slave for the measurement and a master Raspberry PI 3B model responsible for the storage and the communication between the data and the central repository as illustrated in Figure [4]. The data provided will have the following format:

```
<IpAdress | Name | Location | LastMeasurementTime>= <
192.168.0.1|   Rp1   |   Room1   |28-09-2017T11:08:15   ,
192.168.0.2|   Rp2   |   Room2   |28-09-2017T11:08:16   ,
192.168.0.3|   Rp3   |   Room3   |28-09-2017T11:08:17   ,
192.168.0.3|   Rp3   |   Room3   |28-09-2017T11:08:17   ,
… >
```

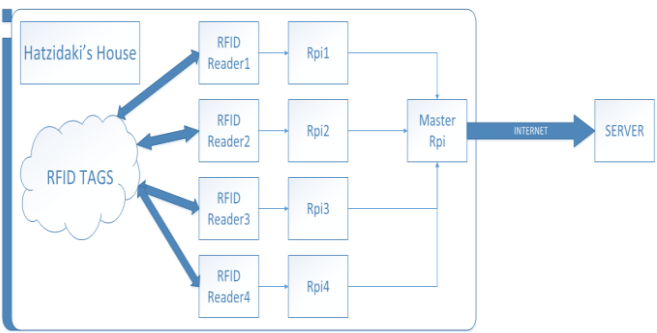

Fig. 4.  Flow chart of the proposed WSN

## IV. WEBSITE TOOLS AND OPTIONS

The programming language, chosen for the development of the above application, was Python, due to its widespread use on the Internet and in particular the Django framework. In the context of the development of the application, the database and the framework of version 1.11 of Django, Python version 3.5 was chosen. The framework that was used is the latest Long-Term Support version that will be supported until April 2020 [8], [9].

The page has four options:

1. *Home*: introductory page.

2. *Sensor Monitoring*: displays statistics for women or men, as well as overall diagrams and conclusions, resulting from the analysis of the data.

3. *Future Plans*: future plans and improvements of the web application.

4. *Contact*: site operator information.

### A. Website Design and Architecture

Initially, we emphasized that the design method of the application was based on the triptych of the projection and the control model as shown in Figure [5]. In particular, the model is a representation of data, at a specific point in time. Emphasis should be placed on the fact that it does not refer to "raw" data, but it creates an interface with them. This level is of utmost importance in the construction of the application, as it provides a subtractive level of base processing, as well as the ability to simultaneously use the same model in multiple databases. Accordingly, a particularly important element is projection, which is a representation of the application. At this level, the computerized system selects the display on the browser, as well as the user interface, in addition to the visualization, providing the user and the data input interface. Finally, there is control over regulating the flow of information between model and projection, through the use of programmed logic for decision making. The options determine the data that will be required by the base and the selected model for this process.

The chosen Django framework defines the third level, in many cases. Therefore, emphasis is given on the remaining two, which are defined in turn, as follows:

1. **_Model:_** access level, validation, relationship processing (primary and foreign keys) and data analysis.

2. **_Template:_** the level of presentation of data and relationships which define the way and volume of displayed information.

3. **_View:_** the level of reference selection and differentiation for interchange between standards.

Finally, in light of the above, it is necessary to highlight the following elements, concerning the architecture of the framework and the development that follows. Initially, there is a clear separation between the stages of the data presentation and the determination of the logic that governs the system. The template is a tool, which is used exclusively to control the display and the presentation of the logic.

For this reason, unlike most applications that involve sites that follow the template "title-header-navigation bar", all items are processed and stored in Django, centrally. In addition, although the results appear to be in HTML, they are in fact more complex entities and they are used to produce text frames or text-based templates, in general. At this point, it must be stressed that disclaiming XML usage for the template, eliminates any human errors and reduces the size of overhead packets. It is also worth mentioning that this tool is characterized by the simplicity of its use, especially we have to process a template in HTML or modify an existing Python code.

Finally, to avoid future mistakes, the framework architecture discourages the use of both variable and complex logical programming structures, in order to avoid writing new commands [10].

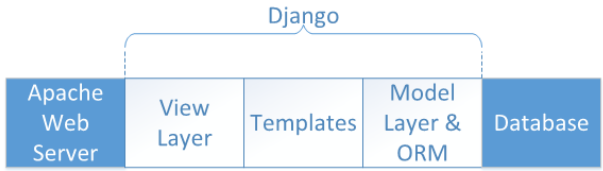

Fig. 5. Architecture design and operations of Django framework

### B. Maintain – Backup – Update the Site contents

In order to ensure the continuous and trouble-free functionality of the site, several security and backup procedures have been included. Firstly, the created site was locally developed on a Linux operating system, where a new local environment was selected, in order to perform all the necessary experimental testings. Afterwards, the final site draft was developed and implemented into a virtual environment at a local host.

Moreover, after each boot of the local network, the first security measure has been the production of a file, which is responsible for the storage of all the necessary settings of the system (requirements.txt). This file basically contained all the packages, variables and files that were associated with the environment and framework's release. The existence of this file is of vital importance to the development of the site and the applications, due to the fact that it enables the user to integrate and control whether new packages can be selected or not. This is an extremely powerful for the operators, aiding them to maintain, upgrade and control the contents of the site, with ease.

In addition, each alternation (removal of an item or the enhancement with a feature) is saved, in accordance with the model-view-control design pattern, in a Python code file (models.py). This file requires that all changes use the migration command and that they are approved by the user, in order to be introduced to the system. Each implementation of this set of commands produces a new Python code file, which will be stored locally and will contain a detailed description of the database structure (scheme) for a predefined time stamp. Additionally, it was selected as the basic means of storage, in order to update the use of the application Git, which is a version control tool, capable of detecting any changes to the system and upgrading our files to the current version, keeping a backup of all the previous settings and files (of the menus and application settings not the data itself). This tool was chosen in order to enable regular backups of the system settings and monitor the changes made, although we have deliberately chosen not to monitor many files. This decision is justified by the need to create a small backup file, as well as to maintain a fast rate of upload/download to the system. For instance, the database is set to be updated at a high rate, automatically, with new sensors measurements, when a specific set of files is executed. As a result, the majority of the system files will not be monitored for backup in the Git respiratory [11]. This feature is derived from the architecture of the Django framework and it is one of its main strengths. As for the data provided by the sensors an open source, self-hosted file sync and share app platform is proposed where the sensors measurements database can be safely stored (nextCloud [12]).

Finally, with regard to the functionality as well as the aesthetic presentation of the site, it was also implemented via Bootstrap 3 [13], a framework which is widely used as an open source tool, for web application development. More specifically, the selection was educated due to the fact that the site had to operate correctly and present its content for a different set of customized applications, running within various browsers, electronic devices of various length screens and user interfaces, which were successfully handled via Bootstrap. The above-mentioned tests were performed through the Chromium program [14] and an early test format of the site is illustrated in Figure [6].

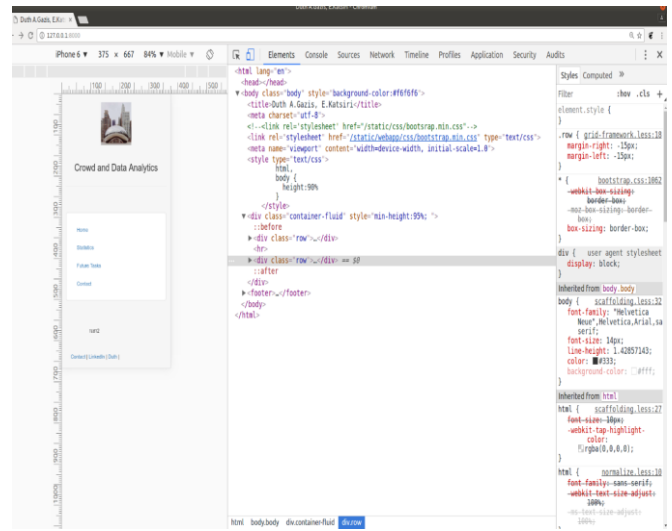

Fig. 6. Testing experiment in order to define the way of the content presentation of the final site for an Iphone 6 device, based on an HTML snippet [15]

### C. Building Evacuation Application

An interesting parameter for extending the usefulness of the site is its use as an intermediate tool for evacuating a building, in case of danger. In particular, the data collected by the sensors placed in each room track the visitors. In addition, there is available through the application a real-time presentation of them as future useful tool. At each time (minute or hour) of the day, the visitor is able to know the number of people who entered and left a certain area (from their id tags). Therefore, in the event of a fire or an earthquake, the application can assist the efforts of the local personnel responsible to help before the arrival of the Fire Service. In this way, it will be a service which will assist in the rapid evacuation of the building for the ground and first floor as illustrated with a red arrow in Figure [7] and [8] respectively.

Moreover, the site may be used in order to send an e-mail, SMS or even a pre-produced telephone alert notification to the respective authorities. Finally, the site itself, since it is running both online and locally, inside the building, can be instantly used, as an escape map. The visitors are able to see a blueprint of the closest escape points, depending on the room they are currently located. Indicatively, schematic illustrations are presented, as following:

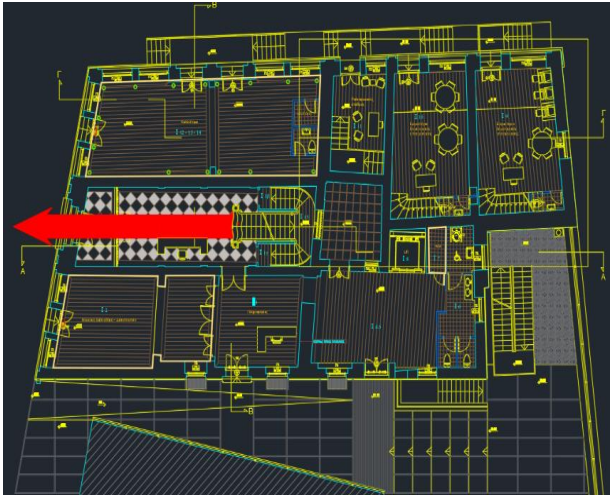

Fig. 7. Evacuation floor plan of the ground floor

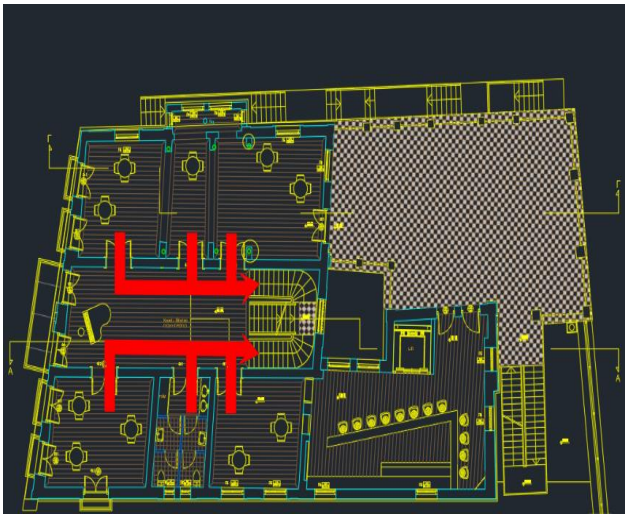

Fig. 8. Evacuation floor plan of the first floor

## V. CONCLUSIONS AND FUTURE WORK

An application was developed in order to count, track and monitor the visitors in a 3 stores high historical building, in Greece. The study was based on the floor plans of the building; therefore, similar implementation may be proposed for other enclosed spaces of the same dimensions. The results of a small set of random generated data produced were particularly positive, hence we propose the extension of this model of study in open space areas, such as musical or sporting events, with a higher crowd density (as long as the point of entrance has specific, constant and preset entry points). The application proposed can be easily implemented with a low power device such as the Raspberry Pi or other low-cost electronic devices because of its minimal computational requirements. The project requires a device with at least 124 [MB] or RAM, 1 [GB] of storage and 512 [MHZ] processor in order to execute the necessary commands in both our front and back end application. Finally, as for the reader a minimum of 4 RFID sensor with several tags is needed and emphasis should only be given in its Bandwidth [16]. The addition of an antenna/s in order to enhance the overall signal is also proposed in order for the coverage to reach each floor to its entity.

As for the context of this publication, both the electrical blueprints and the emergency escape routes were presented. They were mainly consisted of the testing, which was conducted on the ground floor of the building as well as the typical venue space located on the first floor of the establishment. However, we do not exclude the implementation of this model of study on other floors. For instance, the loft which is mainly used for storage, may be used in order to place a local data repository or microcontrollers, so as to accelerate the response of the system.

Moreover, we did not select JavaScript and in particular Node.js [17], in order to produce the web site and the application, despite the fact that they are the most popular web applications, since the design and development did not emphasize the front-end, but the back-end part. Additionally, besides the development of the database's future applications and the extension of the system, we propose the implementation of a system that will include decision-making procedures, depending on the imported data. This will be achieved by various data processing techniques, such as clustering algorithms and Pattern Artificial Neural Networks [18], [19], [20], for which there are plenty of open source libraries, dedicated to this purpose in Python. Additionally, based on the Django framework, it is necessary to extend the middleware and its interface [21] beyond the processes that were automatically generated from it, based on the requests/responses as well as the input/output of the system.

Finally, it is worth mentioning that the system introduced in this publication was created in order to propose a cost-effective method in order to track and monitor the visitors in an area. This method is similar to the ones proposed by the latest technology innovations but, it adds the creation of a system which is efficient, reliable and low cost. Due to the above, it can be selected for a primary application to monitor or a secondary one responsible to evaluate and confirm other existing tracking and monitoring apps (e.g. the ticket sales department, the online reservation application for a concert, etc.)

## ACKNOWLEDGMENTS

The authors wish to thank the Directorate of Technical Works of the Region of Xanthi, East Macedonia-Thrace (ΔΤΕ

ΠΕ ΧΑΝΘΗΣ ΠΑΜΘ) [22] for the dissemination of the floor plans of M. Hatzidakis' residence. Moreover, the authors are grateful to the architect Z. Moschopoulou for her assistance in the analysis and the presentation of the floor plans, as presented in this publication.